\documentclass[11pt,a4paper]{article}
\usepackage{jheppub}
\usepackage{amsmath}
\usepackage{amssymb}

\title{On Non-commutative Corrections of Gravitational Energy in Teleparallel Gravity}
\author[a]{S. C. Ulhoa; }
\affiliation[a]{Faculdade Gama,Universidade de Bras\'{i}lia, Setor Leste (Gama), 72444-240, Bras\'{i}lia-DF, Brazil.}

\author[b,1]{R. G. G. Amorim%
\note{On leave to Universidade de Bras\'{i}lia.}}
\affiliation[b]{Instituto Federal de Goi\'{a}s, Campus Luzi\^{a}nia, 72811-580, Luzi\^{a}nia, GO, Brazil.}
\date{\today}

\emailAdd{sc.ulhoa@gmail.com}
\emailAdd{ronniamorim@gmail.com}

\abstract{In this work we use the theory of Teleparallelism Equivalent to General
Relativity based in non-commutative space-time coordinates. In this context,
we write the corrections of the Schwarzschild solution. As a important result,
we find the corrections of the gravitational energy in the realm of
teleparallel gravity due to the non-commutativity of space-time. Then we
interpret such corrections as a manifestation of quantum theory in gravitational field.}


\begin{document}

\maketitle
\section{Introduction}
\noindent

The notion of non-commutative space coordinates arose with Heisenberg, who wrote a letter to Peierls, in $1930$, about the existence of a uncertain relation in space-time as a possible way to eliminate the singularities existent on the pontual particles self-energy terms. Based on Heinsenberg advice, Peierls applies those ideas on the analysis of the Landau problem. Since then, Peierls comments about those ideas with Pauli, who includes Oppenheimer in the discussion. Oppenheimer comments with his PhD student Hartrand Snyder about the possibility of non-commutative space coordinates \cite{pauli1, pauli2, jackiw}. In that way, Snyder was the first to develop the idea that the space coordinates could not commutate in too close distances scales. In his work, published in $1947$ \cite{snyder1, snyder2}, Snyder proposed a new approach to space-time, which at highly enough energies, the classical space-time is switched to the notion of space-time fragmented on minimum cell size, where there is no more point idea anymore. So, as direct consequence of the non-commutative feature of the operators correspondent to space-time coordinates results on the impossibility to precisely measure a particle position. From the mathematical viewpoint, the simplest algebra of the operators $\widehat{x}^{\mu}$, that represents the hermitian operators correspondent to space-time coordinates, is given on anti-symmetric tensor constant $\theta^{\mu\nu}$,
\begin{equation*}
[\widehat{x}^{\mu},\widehat{ x}^{\nu}]=i\theta^{\mu\nu}.
\end{equation*}
The relations above imply in these uncertain relations
\begin{equation*}
\Delta \widehat{x}^{\mu}\Delta \widehat{x}^{\nu}\geq \frac{1}{2}|\theta^{\mu\nu}|.
\end{equation*}
These last relations suggest that distances of $\sqrt{|\theta^{\mu\nu}|}$ order, effects of the space-time non-commutative turn to be relevant, showing the end of the classical model of space-time and the beginning of a new geometric structure.

From a historical  viewpoint, in quantum field theory, the non-commutative structure of space-time coordinates appeared at the end of $1940$ decade \cite{snyder1, snyder2}. Initially, the idea was to use the non-commutative feature  of the space-time coordinates to control the ultra-violets divergences in quantum electrodynamic. But, the success obtained by renormalization process led the approach based on the non-commutative geometry to oblivion. However, the interest of the physical community resurfaced with the application of non-commutative geometry in non-abelian theories \cite{chans}, in gravitation \cite{kalau, kastler, connes1}, in  standard model \cite{connes2, varilly1, varilly2} and in the understanding of the quantic Hall effect \cite{belissard}. In the last years, the interest in noncommutative theories was renewed with the discover that the dynamic of open string on the presence of an anti-symmetric field can be described, in certain limits, in terms of gauge theories formulated in non-commutative spaces \cite{witten}. Usually the non-commutativity is introduced by means the use of the Moyal product~\cite{Akofor:2008ae} defined as

\begin{eqnarray}
f(x) \star g(x) &\equiv& \exp
\left(  {i \over 2} \theta^{\mu\nu}  {\partial \over \partial x^\mu}
{\partial \over \partial y^\nu} \right) f(x) g(y) |_{y \rightarrow x} \nonumber\\
&=& f(x) g(x) + {i\over 2} \theta^{\mu \nu}
\partial_\mu f  \partial_\nu  g + {1 \over 2!}  {\left( i \over 2 \right)^2} \theta^{\mu_1\nu_1} \theta^{\mu_2\nu_2}(\partial_{\mu_1}   \partial_{\mu_2} f )(\partial_{\nu_1}   \partial_{\nu_2} g )+ \cdots\nonumber\\\label{starproduct}
\end{eqnarray}
with a constant $\theta^{\mu\nu}$. The procedure used for the construction of field theory in non-commutative coordinates is to write the actions replacing the usual product, in the classical lagrangian density, by the Moyal product. In this way, using the properties of non-commutative algebra, we can see that a free field theory has the action of non-commutative field equal to the action of the field in ordinary coordinates. Moreover, in the presence of interactions, effects of non-commutativity are quite significant \cite{witten}.

An alternative theory of gravitation is the so called teleparallel gravity which was introduced by Einstein, as an attempt to unify gravity and electromagnetic field~\cite{einstein}, and Cartan, who developed the main part of the theory~\cite{Cartan}.
From the dynamics point of view, teleparallel gravity and general
relativity predict the same results. On the other hand, teleparallel
gravity allows for the definition of quantities that are physically
of interest, such as the gravitational energy-momentum and
angular momentum tensors which are well behaved~\cite{Maluf:2002zc} when compared
to attempts made in the context of general relativity~\cite{Komar,ADM}.
In addition, none
of such expressions obtained in the realm of general relativity are
dependant on the reference frame, which is certainly not a
desirable feature for energy, momentum and angular momentum.

The expressions for the energy-momentum and angular momentum of the
gravitational field, in the context of the Teleparallelism
Equivalent to General Relativity (TEGR), are invariant under
transformations of the coordinates of the three-dimensional
spacelike surface, they are also dependent on the frame of
reference, as is to be expected. They have been applied consistently
over the years for many different
systems~\cite{maluf,Maluf:2008yy,Maluf:2005sr,daRochaNeto:2011ir,Ulhoa:2010wv}.
The frame dependence is an expected condition for any expression due
to the field since in special relativity the energy of a particle
for a stationary observer is $m$ (with $c=1$), but it is $\gamma m$
for an observer under a Lorentz transformation boost (here $\gamma$
is the Lorentz factor). There is no reason to abandon this feature
once dealing with the gravitational field, and similar behaviour is
expected for momentum and angular momentum.

Therefore we have two successful theories described above and a natural forward step is combine both of them. Then here our aim is study the Teleparallelism Equivalent to General Relativity in the non-commutative space-time context. In section (\ref{tel}) we introduce the concepts of teleparallel gravity, giving the definition of gravitational energy. In this way, in section (\ref{noncommutative}) we present the corrections of the Schwarzschild solution due to the non-commutativity of tetrad fields and the respective correction in the gravitational energy of the whole space-time. Finally in last section we address our concluding remarks.

\bigskip
Notation: space-time indices $\mu, \nu, ...$ and SO(3,1) indices $a,
b, ...$ run from 0 to 3. Time and space indices are indicated
according to $\mu=0,i,\;\;a=(0),(i)$. The tetrad field is denoted by
$e^a\,_\mu$ and the determinant of the tetrad field is represented
by $e=\det(e^a\,_\mu)$.\par
\bigskip

\section{Teleparallel Gravity}\label{tel}
\noindent

The Teleparallelism Equivalent to General Relativity (TEGR) is constructed out of tetrad fields (instead of a metric
tensor) in the Weitzenb\"{o}ck (or Cartan) space-time, in which it
is possible to have distant (or absolute)
parallelism~\cite{einstein}. The tetrad field and metric tensor are
related by

\begin{eqnarray}
g^{\mu\nu}&=&e^{a\mu}e_{a}\,^{\nu}\,; \nonumber\\
\eta^{ab}&=&e^{a\mu}e^{b}\,_{\mu}\,, \label{1}
\end{eqnarray}
where $\eta^{ab}=diag(-+++)$ is the metric tensor of Minkowski
space-time. Then, the Minkowski space-time metric tensor raises and
lowers tetrad indices, similar to the metric tensor in curved
space-time.

Let us start with a manifold endowed with a Cartan
connection~\cite{Cartan},
$\Gamma_{\mu\lambda\nu}=e^{a}\,_{\mu}\partial_{\lambda}e_{a\nu}$,
which can be written as

\begin{equation}
\Gamma_{\mu \lambda\nu}= {}^0\Gamma_{\mu \lambda\nu}+ K_{\mu
\lambda\nu}\,, \label{2}
\end{equation}
where ${}^0\Gamma_{\mu \lambda\nu}$ are the Christoffel symbols and
$K_{\mu \lambda\nu}$ is given by

\begin{eqnarray}
K_{\mu\lambda\nu}&=&\frac{1}{2}(T_{\lambda\mu\nu}+T_{\nu\lambda\mu}+T_{\mu\lambda\nu})\,.\label{3}
\end{eqnarray}
$K_{\mu\lambda\nu}$ is the contortion tensor defined in terms of the
torsion tensor constructed from the  Cartan connection. The torsion
tensor is $T_{\mu \lambda\nu}=e_{a\mu}T^{a}\,_{\lambda\nu}$, with

\begin{equation}
T^{a}\,_{\lambda\nu}=\partial_{\lambda} e^{a}\,_{\nu}-\partial_{\nu}
e^{a}\,_{\lambda}\,. \label{4}
\end{equation}

The curvature tensor obtained from $\Gamma_{\mu \lambda\nu}$ is
identically zero which, using (\ref{2}), leads to

\begin{equation}
eR(e)\equiv -e({1\over 4}T^{abc}T_{abc}+{1\over
2}T^{abc}T_{bac}-T^aT_a)+2\partial_\mu(eT^\mu)\,,\label{5}
\end{equation}
where $R(e)$ is the scalar curvature of a Riemannian manifold in
terms of the tetrad field and $T^\mu=T^b\,_b\,^\mu$. Since the
divergence term in eq. (\ref{5}) does not contribute with the field
equations, the Teleparallel Lagrangian density is

\begin{eqnarray}
\mathfrak{L}(e_{a\mu})&=& -\kappa\,e\,({1\over 4}T^{abc}T_{abc}+
{1\over 2} T^{abc}T_{bac} -T^aT_a) -\mathfrak{L}_M\nonumber \\
&\equiv&-\kappa\,e \Sigma^{abc}T_{abc} -\mathfrak{L}_M\;, \label{6}
\end{eqnarray}
where $\kappa=1/(16 \pi)$, $\mathfrak{L}_M$ is the Lagrangian
density of matter fields and $\Sigma^{abc}$ is given by

\begin{equation}
\Sigma^{abc}={1\over 4} (T^{abc}+T^{bac}-T^{cab}) +{1\over 2}(
\eta^{ac}T^b-\eta^{ab}T^c)\;, \label{7}
\end{equation}
with $T^a=e^a\,_\mu T^\mu$.It is important to note that the
Einstein-Hilbert Lagrangian density is equivalent to its
teleparallel version given by eq. (\ref{6}). Thus both theories
share the same results concerning dynamics and, up to now,
observational data.

Performing a variational derivative of the Lagrangian density with
respect to $e^{a \mu}$, which are the dynamical variables of the
system, the field equations are

\begin{equation}
e_{a\lambda}e_{b\mu}\partial_\nu(e\Sigma^{b\lambda \nu})-
e(\Sigma^{b \nu}\,_aT_{b\nu \mu}- {1\over
4}e_{a\mu}T_{bcd}\Sigma^{bcd}) \;= {1\over {4\kappa}}eT_{a\mu}\,,
\label{8}
\end{equation}
where $T_{a\mu}=e_{a}\,^{\lambda }T_{\mu
\lambda}=\frac{1}{e}\frac{\delta {\mathcal{L}}_{M}}{\delta e^{a\mu
}}$ is the energy-momentum tensor of matter fields. It is possible
to show by explicit calculations the equivalence of eq. (\ref{8})
and Einstein equations~\cite{maluf:335}.

The field equations can be rewritten as

\begin{equation}
\partial_\nu(e\Sigma^{a\lambda\nu})={1\over {4\kappa}}
e\, e^a\,_\mu( t^{\lambda \mu} + T^{\lambda \mu})\;, \label{10}
\end{equation}
where $t^{\lambda\mu}$ is defined by

\begin{equation}
t^{\lambda \mu}=\kappa(4\Sigma^{bc\lambda}T_{bc}\,^\mu- g^{\lambda
\mu}\Sigma^{bcd}T_{bcd})\,. \label{11}
\end{equation}
Since $\Sigma^{a\lambda\nu}$ is skew-symmetric in the last two
indices, it follows that

\begin{equation}
\partial_\lambda\partial_\nu(e\Sigma^{a\lambda\nu})\equiv0\,.\label{12}
\end{equation}
Thus we get

\begin{equation}
\partial_\lambda(et^{a\lambda}+eT^{a\lambda})=0\,\label{13}
\end{equation}
which yields the continuity equation

$$
{d\over {dt}} \int_V d^3x\,e\,e^a\,_\mu (t^{0\mu} +T^{0\mu})
=-\oint_S dS_j\, \left[e\,e^a\,_\mu (t^{j\mu} +T^{j\mu})\right] \,.
$$
It should be noted that the above expression works as a conservation law for the
sum of energy-momentum tensor of matter fields and for the quantity
$t^{\lambda \mu}$. Thus $t^{\lambda \mu}$ is interpreted as the
energy-momentum tensor of the gravitational field~\cite{maluf2}.
Therefore, one can write the total energy-momentum contained in a three-dimensional
volume $V$ of space as

\begin{equation}
P^a = \int_V d^3x \,e\,e^a\,_\mu(t^{0\mu}+ T^{0\mu})\,. \label{14}
\end{equation}
It is worth to note that the above expression is invariant under
coordinate transformation and transforms like a vector under Lorentz
transformations. Such features are desirable and expected for a true
energy-momentum vector.

\section{Non-commutative Corrections for the Gravitational Energy in Schwarzschild Space-Time}\label{noncommutative}
\noindent

In this section we shall start with Schwarzschild space-time~\cite{Dinverno}, described by the following line element

\begin{equation}
ds^2=-\left(1-\frac{2M}{r}\right)dt^2+\left(1-\frac{2M}{r}\right)^{-1}dr^2+r^2\left(d\theta^2+\sin^2{\theta}d\phi^2\right)\,,
\end{equation}
where $M$ is the mass of font. Then a tetrad field adapted to an observer at rest at spatial infinity, which yields the above metric, is
\begin{small}
\begin{equation}
e^a\,_\mu=\left[ \begin {array}{cccc} \sqrt {-g_{{00}}}&0&0&0 \\\noalign{\medskip}0&\sqrt {g_{{11}}}\sin\theta\cos\phi &\sqrt {g_{{22}}}\cos\theta\cos\phi &-\sqrt {g_{{33}} }\sin\phi\\\noalign{\medskip}0&\sqrt {g_{{11}} }
\sin\theta\sin\phi &\sqrt {g_{{22}}}\cos\theta\sin\phi &\sqrt {g_{{33}} }\cos\phi \\\noalign{\medskip}0&\sqrt {g_{{11}} }
\cos\theta &-\sqrt {g_{{22}}}\sin\theta &0\end {array} \right]\,.
\end{equation}
\end{small}
The non-commutativity is introduced by means of the Moyal product, defined in eq. (\ref{starproduct}), between two tetrad field, given by $$\tilde{g}_{\mu\nu}=\frac{1}{2}\left(e^a\,_\mu\star e_{a\nu}+e^a\,_\nu\star e_{a\mu}\right)\,.$$
Thus the new components of the metric tensor due to the non-commutativity of space-time are written in terms of the old ones, up to second order in $\theta^{\mu\nu}$, as
\begin{small}
\begin{eqnarray}
\tilde{g}_{00}&=&g_{00}\,,\\
\tilde{g}_{11}&=&g_{{11}}+\frac{1}{4}\theta^{{23}}\left[\theta^{{23}}g_{{11}}\cos(2\theta) +\theta^{{13}}\sin(2\theta)\left( {\frac {dg_{{11}}}{dr}}\right)\right] + \frac{1}{8}\left[(\theta^{{13}}\sin\theta)^2+(\theta^{{12}})^{2}\right]\left({\frac {d^{2}g_{{11}}}{d{r}^{2}}}\right),\\
\tilde{g}_{22}&=&g_{{22}}-\frac{1}{4}\theta^{{23}}\left[\theta^{{23}}g_{22}\cos(2\theta) +\theta^{{13}}\sin(2\theta)\left({\frac {dg_{{22}}}{dr}}\right)\right]+\frac{1}{8}\left[(\theta^{{13}}\cos\theta)^2 +(\theta^{{12}})^{2}\right]\left({\frac {d^{2}g_{{22}}}{d{r}^{2}}}\right),\\
\tilde{g}_{33}&=&g_{33}+\frac{1}{8}\,\left(\theta^{{13}}\frac {\partial}{\partial {r}}+\theta^{{23}}{\frac {
\partial}{\partial {\theta}}}\right)^2g_{33}+{\frac {(\theta^{{12}})^{2} }{32g_{{33}}^{2}}}\Biggl[\left( {\frac {
\partial ^{2}g_{{33}} }{\partial {\theta}^{2}}}\right)  \left( {\frac {\partial g_{{33}}}{\partial r}}\right) ^{2}+2\,g_{{33}}\left( {\frac {\partial ^{2}g_{{33}}}{\partial
\theta\partial r}} \right) ^{2}+\nonumber\\
&+&\left( {\frac {\partial g_{{33}}}{\partial \theta}} \right) ^{2}\left({\frac {\partial ^{2}g_{{33}} }{\partial
{r}^{2}}}\right) -2\,g_{{33}} \left( {\frac {
\partial ^{2}g_{{33}}}{\partial {\theta}^{2}}} \right) \left({\frac {\partial ^{2}g_{{33}}}{\partial {r}^{2}}}\right) - 2\, \left( {\frac {\partial g_{{33}} }{\partial \theta}} \right)  \left( {\frac {\partial g_{{33}}}{\partial
r}}  \right) \left({\frac {\partial ^{2}g_{{33}}}{
\partial \theta\partial r}}\right) \Biggr] \,,\\
\tilde{g}_{12}&=&\left({\frac {\sin(2\theta)}{32\,g_{{22}}^{3/2} g_{{11}}^{3/2}}}\right)\Biggl\{-\frac{1}{2}(\theta^{{13}})^{2}  \left[g_{{22}}\left( {\frac {dg_{{11}}}{dr}}  \right) - g_{{11}}\left( {\frac {dg_{{22}}}{dr}}\right)\right]^2 +(\theta^{{13}})^{2}g_{{11}} g_{{22}}\biggl[ g_{{22}}\left( {\frac {d^{2}g_{{11}}}{d{r}^{2}}} \right) +\nonumber\\
&+&g_{{11}}\left( {\frac{d^{2}g_{{22}} }{d{r}^{2}}} \right)\biggr]  -8(\theta^{{23}})^{2} g_{{11}}^{2} g_{{22}}
^{2}+4\,\theta^{{13}}\theta^{{23}}  g_{{11}}g_{{22}} \cot(2\theta)  \left[  {\frac {d }{dr}}\left(g_{{22}}g_{{11}}\right)\right] \Biggr\}
\,,
\end{eqnarray}
\end{small}
\begin{small}
\begin{eqnarray}
\tilde{g}_{13}&=&\left({\frac {\theta^{{12}}}{16g_{{33}}^{3/2} g_{{11}}^{1/2}}}\right)\Biggl\{\theta^{{23}}\Biggl[\cos\theta\left(\,\left( {\frac {dg_{{11}}}{dr}} \right) \left( {\frac {\partial g_{{33}}}{\partial \theta}}\right) g_{{33}} -2g_{{11}} \left( {\frac {\partial ^{2}g_{{33}}}{\partial \theta\partial r}}\right)
g_{{33}}+g_{{11}}\left( {\frac {\partial g_{{33}}}{
\partial \theta}}\right)\left( {\frac {\partial g_{{33}}}{\partial r}}\right)\right)+\nonumber\\
&+&\sin\theta\left(-\frac{1}{2}\left( {\frac {dg_{{11}}}{dr}}\right)\left( {\frac {\partial g_{{33}}}{\partial \theta}}\right)^{2}+g_{{33}}\left( {\frac {dg_{{11}} }{dr}}\right)\left( {\frac {\partial ^{2}g_{{33}}}{\partial {\theta}^{2}}}\right)+2g_{{11}}g_{{33}}\left({\frac {\partial g_{{33}}}{\partial r}}\right)\right)\Biggr]+\nonumber\\
&+&\theta^{{13}}\Biggl[\cos\theta\left(-g_{{33}}\left( {\frac {dg_{{11}}}{dr}}\right) \left({\frac {
\partial g_{{33}} }{\partial r}}\right)+g_{{11}} \left( {\frac {\partial g_{{33}}}{\partial r}} \right) ^{2}-2 g_{{11}}g_{{33}}\left({\frac {\partial ^{2}g_{{33}}}{\partial {r}^{2}}}\right)\right)+\sin\theta\cdot\nonumber\\
&\cdot&\left(-\frac{1}{2} \left( {\frac {dg_{{11}}}{dr}} \right) \left( {\frac {\partial g_{{33}}}{\partial \theta}}\right)\left( {\frac {\partial g_{{33}} }{\partial r}} \right)+g_{{33}}\frac{\partial}{\partial r}\left( {\frac {\partial g_{{33}}}{\partial \theta}}\cdot {
\frac {dg_{{11}} }{d{r}}}\right)-\frac{1}{2}\left(\frac{g_{{33}}}{g_{{11}}}\right)\left( {\frac {\partial g_{{33}}}{\partial \theta}}\right) \left( {\frac {dg_{{11}}}{dr}}\right)^{2}\right)\Biggr]\Biggr\}\,,\\
\tilde{g}_{23}&=&\left({\frac { \theta^{{12}}}{32g_{{33}}^{3/2} g_{{22}}^{3/2}}}\right)\Biggl[4\,\theta^{{13}} g_{{22}}^{2}g_{{33}}\sin\theta \left( {\frac {\partial ^{2}g_{{33}}}{\partial {r}^{2}}}\right) -
\theta^{{13}} g_{{33}}\cos\theta\left( {\frac {\partial g_{{33}}}{\partial \theta}}\right)\left( {\frac {dg_{{22}}}{dr}}\right) ^{2}+\nonumber\\
&+&2\,\theta^{{13}} g_{{33}}g_{{22}}\cos\theta \left( {\frac {\partial g_{{33}} }{\partial \theta}}\right)\left( {\frac {d^{
2}g_{{22}}}{d{r}^{2}}} \right) +4\,\theta^{{23}} g_{{22}}^{2}g_{{33}}\cos\theta  \left( {\frac
{\partial g_{{33}}}{\partial r}}\right) -2\,\theta^{{13}}g_{{22}}^{2}\sin\theta\left( {\frac {\partial g_{{33}}}{\partial r}}\right) ^{2}-\nonumber\\
&-&2\,\theta^{{23}}g_{{22}}^{2}\sin\theta \left( {\frac {\partial g_{{33}}}{\partial \theta}} \right)\left(
{\frac {\partial g_{{33}} }{\partial r}}\right) -2\,\theta^{{23}}g_{{33}}g_{{22}}\sin\theta\left( {\frac {dg_{{22}}}{dr}}\right)  \left( {\frac {\partial g_{{33}}}{\partial \theta}} \right) -\theta^{{23}}g_{{22}}\cos\theta \left( {\frac {dg_{{22}}}{dr}} \right)\cdot\nonumber\\
&\cdot&\left( {\frac {\partial g_{{33}}}{\partial \theta}} \right)^{2}+2\,\theta^{{23}}g_{{33}}g_{{22}}\cos\theta\left( {\frac {dg_{{22}} }{dr}}\right) \left( {\frac {\partial ^{2}g_{{33}} }{\partial {\theta}^{2}}}\right) +2\,\theta^{{13}} g_{
{33}}  g_{{22}}\sin\theta\left( {\frac {dg_{{22}}}{dr}} \right)\left( {\frac {\partial g_{{33}}}{\partial r}}\right) -\nonumber\\ &-&\theta^{{13}} g_{{22}}\cos\theta \left( {\frac {dg_{{22}}}{dr}} \right) \left( {\frac {\partial g_{{33}}}
{\partial \theta}}\right) \left({\frac {\partial g_{{33}}}{\partial r}}\right) +2\,\theta^{{13}}g_{{22}} g_{{33}} \cos\theta\left( {\frac {dg_{{22}}}{dr}} \right) \left( {\frac {\partial^{2}g_{{33}}}{\partial \theta\partial r}} \right) +\nonumber\\
&+&4\,\theta^{{23}} g_{{22}}^{2}g_{{33}}\sin\theta \left( {\frac {\partial ^{2}g_{{33}}}{\partial \theta\partial r}}\right) \Biggr]\,.
\end{eqnarray}
\end{small}
It should be noted that such components are always quadratic in the non-commutative parameter $\theta^{ij}$. This indicates that the corrections have very tiny values. Therefore the new metric tensor obeys approximately Einstein's equations. We point out that the corrections above are different from the ones presented in ref. \cite{Chaichian2008573}.

Now we have to write another tetrad field for the new components of the metric tensor, but still adapted to a rest frame, which is the referential frame we would like to analyze the problem of the gravitational energy. Hence $\tilde{g}_{\mu\nu}=\tilde{e}^{a}\,_{\mu}\tilde{e}_{a\nu}$, where $\tilde{e}^{a}\,_{\mu}$ is given by
\begin{small}
\begin{equation}
\tilde{e}^a\,_\mu=\left[ \begin {array}{cccc} A&0&0&0\\
\noalign{\medskip} 0&\scriptstyle B\sin\theta\cos\phi +Er\cos\theta\cos\phi -F
r\sin\theta\sin\phi &\scriptstyle Cr\cos\theta\cos\phi -Gr\sin\theta\sin\phi&\scriptstyle -Hr\sin\theta\sin\phi \\
\noalign{\medskip}0&\scriptstyle B\sin\theta\sin\phi +Er\cos\theta\sin\phi +Fr\sin\theta\cos\phi & \scriptstyle Cr\cos\theta\sin\phi+Gr\sin\theta\cos\phi & \scriptstyle Hr\sin\theta\cos\phi\\
\noalign{\medskip}0& \scriptstyle B\cos\theta -Er\sin\theta&\scriptstyle -Cr\sin\theta&0\end {array} \right],
\end{equation}
\end{small}
where
\begin{eqnarray}
A&=&\sqrt {-\tilde{g}_{{00}} }\,,\nonumber\\
B&=&\frac{\delta_1}{\delta}\,,
\nonumber\\
C&=&{\frac {\delta}{r\sqrt {\tilde{g}_{{33}}}}}\,,
\nonumber\\
E&=&{\frac {\tilde{g}_{{12}} \tilde{g}_{{33}} -\tilde{g}_{{23}} \tilde{g}_{{13}}}{r\sqrt {\tilde{g}_{{33}}}\,\delta}}\,,
\nonumber\\
F&=&{\frac {\tilde{g}_{{13}}}{\sqrt {\tilde{g}_{{33}}}\,r\sin\theta}}\nonumber\\
G&=&{\frac {\tilde{g}_{{23}}}{\sqrt {\tilde{g}_{{33}}}\,r\sin\theta}}\,,
\nonumber\\
H&=&{\frac {\sqrt {\tilde{g}_{{33}}}}{r\sin\theta}}\,.
\end{eqnarray}
The quantities $\delta$ and $\delta_1$ are defined by

\begin{equation*}
\delta^2=\tilde{g}_{22}\tilde{g}_{33}-\tilde{g}_{23}^2
\end{equation*}
and
\begin{equation*}
\delta_1^2=\tilde{g}_{11}\delta^2-\tilde{g}_{22}\tilde{g}_{13}^2-\tilde{g}_{33}\tilde{g}_{12}^2+2\tilde{g}_{12}\tilde{g}_{23}\tilde{g}_{13}\,.
\end{equation*} In order to compute the gravitational energy, the relevant non-vanishing components of torsion tensor are

\begin{eqnarray}
\tilde{T}_{112}&=&-\frac{1}{2}\,\left(\frac{\partial \tilde{g}_{11}}{\partial\theta}\right)+\frac{1}{2\delta^2}\left[\alpha\left(\frac{\partial \tilde{g}_{22}}{\partial r}\right)+\beta\left(\frac{\tilde{g}_{23}}{\tilde{g}_{33}}\right)\left(\frac{\partial \tilde{g}_{33}}{\partial r}\right)-2\beta\left(\frac{\partial \tilde{g}_{23}}{\partial r}\right)\right]\,,\nonumber\\
\tilde{T}_{212}&=&\frac{1}{2}\left(\frac{\partial \tilde{g}_{22}}{\partial r}\right)-\left(\frac{\partial \tilde{g}_{12}}{\partial\theta}\right)-\delta_1\tilde{g}_{33}^{-1/2}+\frac{1}{2\delta^2}\left[\alpha\left(\frac{\partial \tilde{g}_{22}}{\partial \theta}\right)+\beta\left(\frac{\tilde{g}_{23}}{\tilde{g}_{33}}\right)\left(\frac{\partial \tilde{g}_{33}}{\partial \theta}\right)-2\beta\left(\frac{\partial \tilde{g}_{22}}{\partial \theta}\right)\right]\,,\nonumber\\
\tilde{T}_{312}&=&\frac{1}{2\tilde{g}_{33}}\,\left[2\tilde{g}_{33}\left(\frac{\partial \tilde{g}_{23}}{\partial r}\right)-\tilde{g}_{23}\left(\frac{\partial \tilde{g}_{33}}{\partial r}\right)+\tilde{g}_{13}\left(\frac{\partial \tilde{g}_{33}}{\partial\theta}\right)-2\tilde{g}_{33}\left(\frac{\partial \tilde{g}_{13}}{\partial \theta}\right)\right]\,,\nonumber\\
\tilde{T}_{113}&=&\frac{1}{2}\,\left(\frac{\tilde{g}_{13}}{\tilde{g}_{33}}\right)\left(\frac{\partial \tilde{g}_{33}}{\partial r}\right)\,,\nonumber\\
\tilde{T}_{213}&=&-\frac{1}{2\tilde{g}_{33}\delta}\left[2\tilde{g}_{33}\beta\cos\theta+2\tilde{g}_{23}\tilde{g}_{33}^{1/2}-\tilde{g}_{23}\delta\left(\frac{\partial \tilde{g}_{33}}{\partial r}\right)\right]\,,\nonumber\\
\tilde{T}_{313}&=&\frac{1}{2}\,\left(\frac{\partial \tilde{g}_{33}}{\partial r}\right)-\frac{1}{\delta}\left(\alpha\cos\theta+\delta_1\tilde{g}_{33}^{1/2}\sin\theta\right)\,,\nonumber\\
\tilde{T}_{123}&=&-\frac{1}{2}\,\left(\frac{\tilde{g}_{13}}{\tilde{g}_{33}}\right)\left(\frac{\partial \tilde{g}_{33}}{\partial \theta}\right)-\frac{1}{\delta}\left(\beta\cos\theta+\delta_1\tilde{g}_{23}\tilde{g}_{33}^{-1/2}\sin\theta\right)\,,\nonumber\\
\tilde{T}_{223}&=&\frac{1}{2}\,\left(\frac{\tilde{g}_{23}}{\tilde{g}_{33}}\right)\left(\frac{\partial \tilde{g}_{33}}{\partial \theta}\right)\,,\nonumber\\
\tilde{T}_{323}&=&\frac{1}{2}\,\left(\frac{\partial \tilde{g}_{33}}{\partial\theta}\right)-\delta\cos\theta\,.
\end{eqnarray}
where
$$\alpha=\tilde{g}_{33}\tilde{g}_{12}-\tilde{g}_{23}\tilde{g}_{13}$$ and $$\beta=\tilde{g}_{12}\tilde{g}_{23}-\tilde{g}_{22}\tilde{g}_{13}\,.$$
Using eq. (\ref{7}), it yields

\begin{eqnarray}
\tilde{\Sigma}^{001}&=&\frac{1}{2(-\tilde{g}_{00})\delta_1^2}\biggl[-\tilde{T}_{112}\left(\delta^2\tilde{g}_{12}+
\alpha\tilde{g}_{11}\right)-\tilde{T}_{212}\left(\delta^2\tilde{g}_{22}+\alpha\tilde{g}_{12}\right)-
\tilde{T}_{312}\left(\delta^2\tilde{g}_{23}+\alpha\tilde{g}_{12}\right)-\nonumber\\
&-&\tilde{T}_{113}\left(\delta^2\tilde{g}_{13}-\beta\tilde{g}_{11}\right)-\tilde{T}_{213}
\left(\delta^2\tilde{g}_{23}-\beta\tilde{g}_{12}\right)-\tilde{T}_{313}\left(\delta^2\tilde{g}_{33}-
\beta\tilde{g}_{13}\right)+\tilde{T}_{123}(\alpha\tilde{g}_{13}+\nonumber\\
&+&\beta\tilde{g}_{12})+\tilde{T}_{223}\left(\alpha\tilde{g}_{23}+\beta\tilde{g}_{22}\right)+\tilde{T}_{323}\left(\alpha\tilde{g}_{33}
+\beta\tilde{g}_{23}\right)\biggr]\,,
\end{eqnarray}
then it is possible to find $\tilde{\Sigma}^{(0)01}=\tilde{e}^{(0)}\,_0\tilde{\Sigma}^{001}$, which after some algebraic manipulations is written as
\begin{small}
\begin{eqnarray}
4e\,\tilde{\Sigma}^{(0)01}&=&\left(\frac{2}{\delta_1}\right)\left\{\tilde{g}_{33}^{1/2}\delta_1+\left(\frac{\delta\delta_1}{\sqrt{\tilde{g}_{33}}}\right)\sin\theta
-\frac{1}{2}\frac{\partial }{\partial r}\left(\tilde{g}_{33}\tilde{g}_{22}\right)+\frac{\partial }{\partial \theta}\left(\tilde{g}_{33}\tilde{g}_{12}\right)+\tilde{g}_{23}\left[\left(\frac{\partial \tilde{g}_{23}}{\partial r}\right)-\left(\frac{\partial \tilde{g}_{13}}{\partial \theta}\right)\right]\right\}+\nonumber\\
&+&\left(\frac{1}{\delta^2\delta_1}\right)\left\{\left(\tilde{g}_{23}\tilde{g}_{13}-\tilde{g}_{12}\tilde{g}_{33}\right)\left[\frac{\partial\left( \tilde{g}_{22}\tilde{g}_{33}\right)}{\partial \theta}\right]+2\tilde{g}_{33}\left(\tilde{g}_{12}\tilde{g}_{23}-\tilde{g}_{22}\tilde{g}_{13}\right)\left(\frac{\partial \tilde{g}_{23}}{\partial \theta}\right)\right\}\,.
\end{eqnarray}
\end{small}
Then taking the limit $r\rightarrow\infty$ it yields

\begin{small}
\begin{equation}
4e\,\tilde{\Sigma}^{(0)01}=4\,M\sin\theta-(\theta^{23})^2\,\left(\frac{\cos^2\theta}{\sin\theta}\right)\left[ M\left(\frac{11}{8}+\cos^2\theta\right)+\lim_{r\rightarrow\infty}\,\frac{3}{2}
r\left(\frac{1}{4}+\cos^2\theta\right)\right]\,.
\end{equation}
\end{small}The energy is calculated using eq. (\ref{14}) and performing a regularization procedure in order to avoid divergences. Hence we get

\begin{equation}
\tilde{P}^{(0)}= M+\frac{1}{4}\,\left(\frac{49}{3}-\frac{15}{2}\ln{2}\right)\,(\theta^{23})^2\,M\,,
\end{equation}
thus the correction in the gravitational energy due to the non-commutativity of space-time is $$\Delta P^{(0)}=\frac{1}{4}\,\left(\frac{49}{3}-\frac{15}{2}\ln{2}\right)\,(\theta^{23})^2\,M\,.$$ Alternatively the need for a regularization procedure could be contoured by means the choice $\theta^{23}=0$, in such a way there would be no correction for the Schwarzschild energy.

\section{Conclusion}
\noindent

In this work we start with Schwarzschild space-time, then we give the corrections due to the non-commutativity of space-time. Here it is introduced by replacing the normal product between tetrads by the Moyal product, rather than applying such a procedure in lagrangian density. The new metric tensor leads to a new tetrad field which is used to calculate the gravitational energy of space-time. It is well known that the energy of Schwarzschild space-time is equal to $M$, therefore we get a correction in the energy equal to $\Delta P^{(0)}$. Since the non-commutative parameter is arbitrary (it should be given by experimental data) we speculate that such a correction in the gravitational energy can be associated to quantum effects in the realm of gravitational field. If the correction represents the energy of gravitons, then it should be proportional to the Planck's constant. On the other hand the correction is proportional to the mass of the font, which could mean a new kind of quantization associated to the mass of a black-hole or a star, for example. This has been expressed in ref. \cite{Ulhoa:2011py}. For future works we intend to investigate the corrections of the gravitational energy in the context of Kerr space-time on the outer event horizon. We also want to study the solutions of the non-commutative equations that come from the lagrangian density replaced by the Moyal product.

\bibliographystyle{JHEP}
\bibliography{ref}

\end{document}